# Monolithic Germanium–Tin on Si Avalanche Photodiodes

Justin Rudie, Sylvester Amoah, Xiaoxin Wang, Rajesh Kumar, Grey Abernathy, Steven Akwabli, Perry C. Grant, Jifeng Liu, *Senior Member, IEEE*, Baohua Li, Wei Du, *Senior Member, IEEE*, Shui-Qing Yu, *Senior Member, IEEE*

*Abstract*—We demonstrate monolithically grown germanium-tin (GeSn) on silicon avalanche photodiodes (APDs) for infrared light detection. A relatively thinner Ge buffer design was adopted to allow effective photo carriers to transport from the GeSn absorber to the Si multiplication layer such that clear punch-through behavior and a saturated primary responsivity of 0.3 A/W at 1550 nm were observed before avalanche breakdown in GeSn/Si APDs for the first time. The spectral response covers 1500 to 1700 nm. The measured punch-through and breakdown voltages are 15 and 17 V, respectively. Undisputed multiplication gain was obtained with the maximum value of 4.5 at 77 K, and 1.4 at 250 K, directly in reference to the saturated primary responsivity from the same device rather than a different GeSn p-i-n photodiode in previous reports. A peak responsivity was measured as 1.12 A/W at 1550 nm and 77 K.

Manuscript received xxxx xx, 2024; revised xxxx xx, 2024; accepted xxxx xx, 2024. Date of publication xxxx xx, 2024; date of current version xxxx xx, 2024. This work was supported by the Air Force Research Laboratory (AFRL)/AFWERX (Contract No. FA864922P0744). Justin Rudie and Sylvester Amoah contributed equally to the work. *(Corresponding authors: Xiaoxin Wang, Wei Du, and Shui-Qing Yu).*

Justin Rudie is with Material Science and Engineering, University of Arkansas, Fayetteville, AR 72701 USA and Department of Electrical Engineering and Computer Science, University of Arkansas, Fayetteville, AR 72701 USA (e-mail: jrudie@uark.edu).

Sylvester Amoah, Steven Akwabli and Rajesh Kumar are with Department of Electrical Engineering and Computer Science, University of Arkansas, Fayetteville, AR 72701 USA (e-mails: samoah@uark.edu; sakwabli@uark.edu; rajeshk@uark.edu).

Xiaoxin Wang and Jifeng Liu are with Thayer School of Engineering, Dartmouth College, Hanover, NH 03755 USA (e-mails: jifeng.liu@dartmouth.edu; xiaoxin.wang@dartmouth.edu).

Grey Abernathy is with Arktonics, LLC, 1339 S. Pinnacle Dr., Fayetteville, AR, 72701, USA, Department of Electrical Engineering and Computer Science, University of Arkansas, Fayetteville, AR 72701 USA and Materials Science & Engineering Program, University of Arkansas, Fayetteville, AR, 72701, USA (e-mail: grey.abernathy@arktonics.com).

Perry C. Grant and Baohua Li are with Arktonics, LLC, 1339 S. Pinnacle Dr., Fayetteville, AR 72701 USA (e-mails: Perry.Grant@arktonics.com; baohua.li@arktonics.com).

Wei Du and Shui-Qing Yu are with the Department of Electrical Engineering and Computer Science, University of Arkansas, Fayetteville, AR 72701 USA and the Institute for Nanoscience and Engineering, University of Arkansas, Fayetteville, AR 72701 USA (e-mails: weidu@uark.edu; syu@uark.edu).

Color versions of one or more of the figures in this article are available online at http://ieeexplore.ieee.org

## I. INTRODUCTION

AVALANCHE photodiodes (APDs) operating in infrared (IR) range have been widely used in various applications where high sensitivities are needed, including fiber-optic communications, light detection and ranging (LiDAR) systems, military activities, medical sensing, etc. [1-3]. Compared with standard p-i-n based photodiode, APDs can offer higher sensitivity due to internal gain enabled by avalanche multiplication. Traditional III-V compounds and mercury cadmium telluride (HgCdTe) based technologies could achieve infrared detection at the costs of high material expense and especially the relatively large excess noise factor [4-7]. Many research efforts are now focusing on alternative multiplication materials such as Si which has low impact ionization ratio $k$ and thus low excess noise factor [8], [9]. However, Si based APDs cannot operate in broad IR wavelength range. A remarkable breakthrough has been accomplished by the demonstration of separate absorption-charge-multiplication (SACM) structure applying to Ge on Si APDs, where the Ge absorber and Si multiplication layer are spatially separated and they are connected by a Si charge layer to balance the electric field profile within the device [10-12]. Due to the bandgap of Ge, this type of APD targets 1550 nm operation and the cutoff wavelength is ~1600 nm.

For longer wavelength detection, III-V materials are commonly used but a major challenge lies in inserting III–V materials into a silicon complementary metal oxide semiconductor (CMOS) fabrication facility. Therefore, developing an IR APD based on a monolithic and CMOS compatible process is highly desirable. Recent study of the group IV alloy GeSn holds great promise for high performance IR photodetectors [13-21]. By adjusting Sn composition, GeSn alloy could cover shortwave-IR (SWIR) and mid-IR (MIR) ranges. GeSn p-i-n photodiodes have demonstrated the spectral response extending to beyond 3.0 μm, and low temperature specific detectivity (D*) is only one order of magnitude lower than that of commercial extended InGaAs detector operating in the same wavelength [16].

Some preliminary results for GeSn APD development have been reported [22-25]. In some work a relatively thick Ge buffer layer was used, which may prohibit the collection of photo carriers from the GeSn absorber and therefore avalanche multiplication is very difficult to achieve. A GeSn in Si APD without Ge buffer was presented as well. The relatively large lattice mismatch between Si and GeSn would lead to high defect density at Si/GeSn interface, resulting in photo carriers becoming trapped at the interface which consequently may fail the device. Notably, none of the previous reports on GeSn/(Ge buffer)/Si APDs have shown clear punch-through behavior or saturation in responsivity before avalanche breakdown, making it impossible to accurately determine the primary responsivity and avalanche gain in reference to the same device. While several papers claimed that GeSn p-i-n diodes were used as references to determine the multiplication gain, the data of these reference devices were not presented, leaving some ambiguity in the results. In this work, a GeSn on Si APD utilizing SACM structure was demonstrated. A relatively thin Ge buffer of 100 nm was adopted to facilitate carrier transportation and demonstrate clear punch-through behavior as well as photoresponse saturation before avalanche breakdown in GeSn/Ge(buffer)/Si APDs for the first time. The distinct punch-through and breakdown voltages were measured as -15 and -17 V, respectively. The avalanche gain was measured in reference to the saturated primary responsivity of the same device before avalanche breakdown, offering accurate and undisputed multiplication gain data without any complication from using another p-i-n diode as the external reference. The internal gain was obtained even at 250 K (~1.4). At 77 K, the maximum gain of 4.5 was achieved with a peak responsivity of 1.12 A/W at 1550 nm.

## II. DESIGN PRINCIPLE BASED ON DEVICE MODELING

The schematic SACM structure of a surface illuminated GeSn/Si APD is shown in Fig. 1(a). The SACM structure can leverage the high detection in the infrared range of GeSn absorber and low excess noise factor of Si multiplication layer with impact ionization ratio $k \sim 0.02$ [26]. Since there is a relatively large lattice mismatch between GeSn and Si, normally a thick Ge buffer (> 700 nm) is often required for high quality epitaxial GeSn layer. Indeed, this strategy was adopted in previous work on GeSn/Si APDs. However, a thick Ge buffer would dramatically reduce collection efficiency of photo carriers from the front-end GeSn absorber to the Si multiplication layer, leading to the failure of GeSn/Si SACM APDs. Moreover, it is well acknowledged that a good APD should have a punch-through voltage less than the breakdown voltage, and a broad plateau range is preferred in the I-V curve to determine the punch-through voltage [27]. This is especially important for high-speed APDs since sufficient electric field is extended to the GeSn absorber for fast photo carrier drift transport only after the punch-through is reached. If an APD reaches breakdown voltage before punch-through, the bandwidth and primary responsivity will be severely limited by the slow carrier diffusion in the GeSn region due to the lack of electric field.

However, a notable issue is that punch-through behavior has not yet been demonstrated in GeSn/(Ge buffer)/Si APD devices reported so far. To understand the underlying reason, we modeled room-temperature dark current density as a function of bias voltage (J-V characteristics) of exemplary $Ge_{0.9}Sn_{0.1}$/Ge buffer/Si APDs with different Ge buffer layer thicknesses in Fig. 1(b). Without Ge buffer, punch-through and breakdown voltages can be seen at ~8 V and ~25 V, respectively. This clear punch-through before avalanche breakdown is desirable for APD devices. Nevertheless, the defective GeSn/Si interface due to large lattice mismatch will trap majority photo carriers, resulting in very few carriers reaching the Si multiplication layer, and eventually cause device failure. On the other hand, increasing the Ge buffer thickness would raise the punch-through voltage, narrow the plateau range and even result in breakdown occurring before the punch-through for 700 nm thick Ge buffer based on Fig. 1(b), an undesirable result for APD operation. The dependence of internal avalanche gain on Ge buffer thickness is further plotted in Fig. 1(c). As Ge buffer thickness increases, the unity gain (i.e. primary responsivity) occurs at higher voltage. For 700 nm buffer, no gain was obtained before breakdown. In contrast, for 100 nm buffer, the unity gain was observed at -17 V, indicating a rational design. Figure 1(d) further shows electric filed distribution at -25 V influenced by Ge buffer thickness. The colored shaded area outlines the device layer stack using 100-nm-thick Ge buffer. Thicker Ge buffer leads to lower electric filed in the GeSn absorber, which is unfavorable to transport the carriers. The thinner buffer (less than 100 nm) has an electric field strong enough to sweep the photo generated electrons in GeSn absorber to the Si multiplication layer.

Therefore, the modeling clearly shows that, very different from conventional design of Ge buffers for p-i-n diodes, a thinner Ge buffer design is the basic principle for high performance GeSn/Ge/Si APDs to achieve punch-through well before avalanche breakdown is reached. Considering the trade-off between low dark current density and punch-through, the Ge buffer thickness was designed as 100 nm in this work, corresponding to a punch-through of ~ -15 V. This design also offers notable avalanche gain at relatively low reserve bias for up to 10 at. % Sn composition in response to optical excitation at $\lambda$=2 μm (Fig. 1(c)). By developing a unique deposition technique, a high-quality GeSn layer can be grown on this thin Ge buffer.

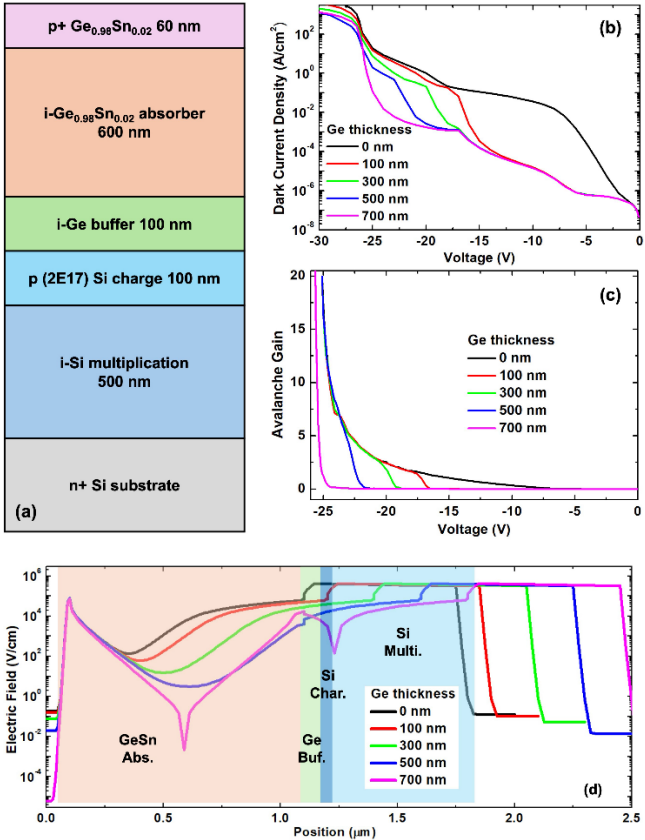

**Fig. 1.** **(a)** Schematic SACM structure showing the design of $Ge_{1-x}Sn_x$/Ge buffer/Si APDs. **(b)** Theoretically modelled dark J-V curves for different Ge buffer thicknesses at 10% Sn composition at 300 K. **(c)** Internal avalanche gain vs. reverse bias voltage for different Ge buffer thicknesses under optical excitation at λ=2 μm for 10% Sn composition. **(d)** Electric field distribution at -25 V for different Ge buffer layer thickness for 10% Sn composition.

The GeSn absorber is designed to be relatively thick (600 nm) to have sufficient light absorption. A 500-nm-thick intrinsic Si multiplication layer was found to be sufficiently thick to offer notable avalanche gain through the modeling in Fig. 1(c), and a 100-nm-thick Si charge layer with p-type doping concentration of $2×10^{17}$ cm$^{-3}$ is adopted in this work, as widely used in Ge/Si APDs. In the experimental work, we started with a relatively lower Sn composition of 2 at. % to demonstrate the design principle of thin Ge buffer layers while keeping a high material quality in the GeSn absorber region.

### III. MATERIAL GROWTH AND CHARACTERIZATION

The GeSn/Si APD layer stack was epitaxially grown using an industrial standard ASM Epsilon® 2000 Plus reduced pressure chemical vapor deposition (RPCVD) reactor with commercially available SiH₄, GeH₄, and SnCl₄ as precursors. The key step is the growth of 100-nm-thick Ge buffer. A low temperature growth approach was applied. Subsequently, a 600-nm-thick undoped GeSn absorber with 2% Sn composition was grown and was capped with a 60-nm-thick p+ doped GeSn (also 2% Sn) as contact layer.

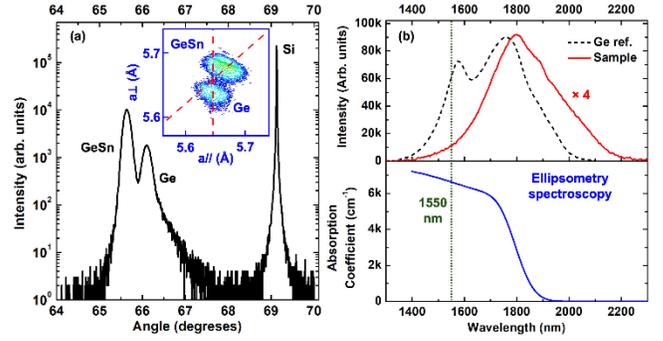

**Fig. 2.** **(a)** XRD showing clearly resolved Ge and GeSn peaks. Inset: RSM contour plot showing an almost relaxed GeSn layer. **(b)** PL and ellipsometry spectroscopies showing emission from GeSn layer and absorption edge.

X-ray diffraction (XRD) measurement was carried out using a Panalytical X'Pert Pro MRD diffractometer equipped with a 1.8 kW Cu Kα1 X-ray tube, a standard four-bounce Ge (220) monochromator, and a Pixel detector. Figure 2(a) shows the 2θ-Ω scan. The Ge buffer and GeSn peaks are clearly observed, indicating the high material quality. The reciprocal space mapping (RSM) plotted in inset shows that the Ge buffer is with 0.16% tensile strain while the GeSn is almost relaxed with a residual compressive strain of 0.04%.

Photoluminescence (PL) spectroscopy was performed using a standard off-axis configuration with a lock-in amplifier. A 1064 nm nano-second pulsed laser was used as the excitation source. The PL emission was collected by a spectrometer and then sent to an InSb detector with a cutoff wavelength at 5.0 μm. Figure 2(b) top image shows PL spectrum at room temperature. The PL of Ge reference was also shown for comparison. The obvious peak red shift to 1800 nm indicates that the emission is from GeSn absorber. Ellipsometry spectroscopy was plotted in Fig. 2(b) bottom, exhibiting the absorption edge at ~1850 nm. The optical characterization results agree well with theoretical study of bandgap energy for 2% Sn alloy. Note that at 1550 nm, the absorption coefficient is greater than 6,000 cm$^{-1}$, indicating an excellent absorber with good material quality.

## IV. DEVICE CHARACTERIZATION RESULTS AND DISCUSSIONS

The sample was fabricated into APD devices with square mesa and side lengths of 500 and 1000 μm. Standard photolithography and wet chemical etching processes were applied to define the patterns, followed by metallization process with top contact on the mesa surface and bottom contact at the backside of Si substrate. The 10/300 nm Cr/Au was used to form ohmic contact.

### A. Electrical Characterization

Dark current-voltage (I-V) characterization was performed using a Keysight B2902b source measurement unit (SMU). For photocurrent measurements the optical excitation was provided by a 1.55 μm laser with a power of 0.5 mW. For low temperature characterization, devices were wire bonded and mounted in an electrically isolated cryostat with CaF₂ windows to maximize IR light transmission.

Figure 3(a) shows dark current of a 500 μm device at different temperatures. The APD shows typical rectifying behavior at all temperatures. The dark current decreases with decreasing temperature which is consistent with the generation–recombination model. Notably, the dark current starts to increase steeply at ~ -12V, indicating the onset of the p-Si charge layer depletion that enables transport of carriers generated in the GeSn region to the Si multiplication layer. The dashed curve is the simulated room temperature I-V, which agrees well with measured I-V curve. The simulation used a 100 nm-thick p-Si charge layer with $1.4 \times 10^{17}$ cm⁻³ hole concentration and a 400-nm-thick intrinsic Si multiplication layer, both very close to the designed values within the fabrication error range.

Figure 3(b) shows comparison of dark and photo currents at 77 K. There are three regions for increased photo current. The first region (0 to -5 V) might be due to the heterojunction barriers, including GeSn/Ge junction and Ge/Si junction; the second region (-5 to -15 V) is because of the enhanced carrier collection efficiency as a results of depletion region gradually expanding towards the GeSn absorber, until the punch-through that ideally all photo carriers can be collected; the third region (after -15 V) is attributed to the multiplication gain that makes the current increase rapidly. The photo currents at 150 K and 250 K are plotted in Fig. 4(c) and (d), respectively. A similar behavior was observed in 150 K photo current. At 250 K, the difference between photo and dark current is still distinct, indicating excellent device quality.

The punch-through voltage at which point the depletion region penetrates into the GeSn absorber is hard to determine exactly from the I-V characteristics alone. The punch-through voltage can be determined by capacitance-voltage (C-V) measurement (shown in Figure 4) in comparison with the photocurrent-voltage results in Fig. 3(b)-(d) as well as Fig. 6. At -15 V, the device capacitance shows a steep decrease, a characteristic feature of punch-through [28]. Correspondingly, the photocurrent/responsivity reaches a plateau at -15 V, indicating that all photo carriers generated in GeSn were transferred to the Si multiplication layer. This result further confirms that the punch-through voltage is -15 V. The breakdown voltage is defined as the bias point at which the current reaches 1 mA. At 77 K the breakdown voltage was extracted as -17 V.

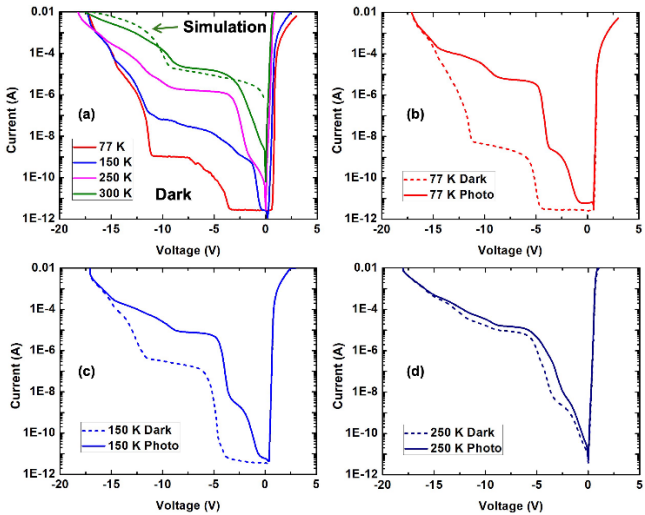

**Fig. 3**. **(a)** Dark current at different temperatures. Dashed curve is simulation of dark current at 300 K. **(b)-(d)** Photo current (solid curves) compared with dark current (dashed curves) at 77 K, 150 K, and 250 K.

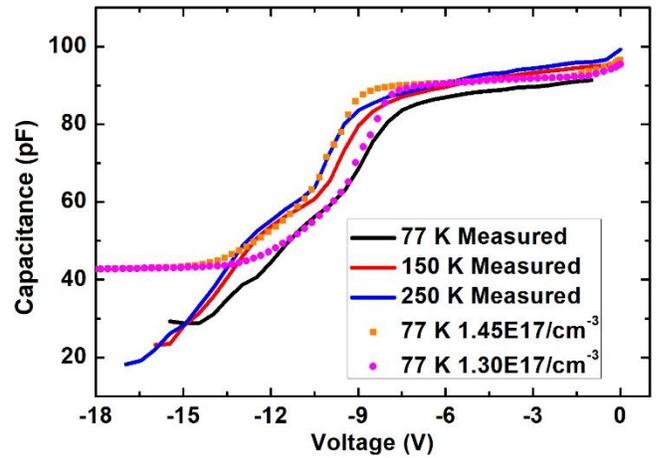

**Fig. 4.** C-V characteristics at different temperatures. Solid curves are measured data and solid dots are simulations with two doping concentrations in Si charge layer.

Temperature dependent C-V measurements were further performed using a Keithley 236 capacitance measurement unit. The measured curves are plotted in Fig. 4, showing that the characteristics are similar at different temperatures. All curves exhibit a plateau feature around -15 V, clearly indicating the punch-through has been achieved. The larger capacitance at higher temperatures is likely due to a higher density of residual carriers in GeSn due to defects. C-V modeling data at 77 K are also plotted (solid dots). When a lower doping concentration of Si charger was used, the simulated curve matches well with measured data. This might be due to the fact that at 77 K the dopants in charge layer are partially activated, leading to a lower effective doping centration.

### B. Gain and Responsivity Characterization

Spectral response was performed using a Fourier transform infrared spectroscope (FTIR) equipped with an internal IR light source. Photo response was measured across a 47 kΩ resister in series with the device under test (DUT) using a pre-amplifier that fed into the FTIR. A 1300 nm long pass filter was used to block the absorption in the Si multiplication layer. For responsivity measurement, the same 1.55 μm laser and 0.5 mW power utilized in photo response measurements was used. The signal was chopped at a frequency of 377 Hz and measured across a series resistor of 15 Ω, then fed into a lock in amplifier. Responsivity was calculated using photocurrent flowing through the resistor. Moreover, the APD gain was obtained by normalizing corresponding responsivity to the primary responsivity.

Figure 5 shows temperature-dependent spectral response. The cut-off wavelength moves towards longer wavelength as temperature increases, from 1700 nm to 1850 nm corresponding to 77 K and 250 K, respectively. The longer wavelength cut-off compared to Ge indicates that the photo response is originally from GeSn layer. The measured spectral response matches well with ellipsometry spectroscopy in Fig. 2.

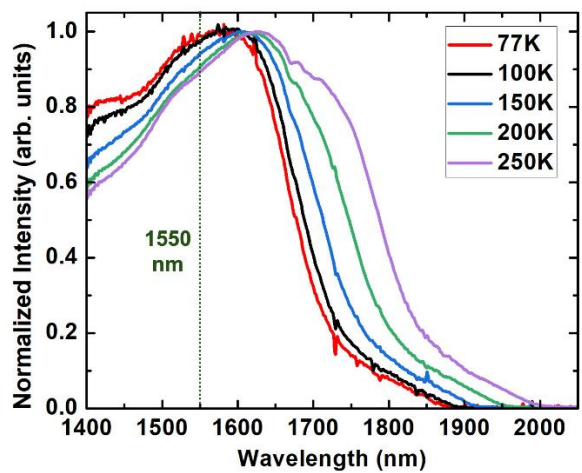

**Fig. 5.** Temperature-dependent spectral response.

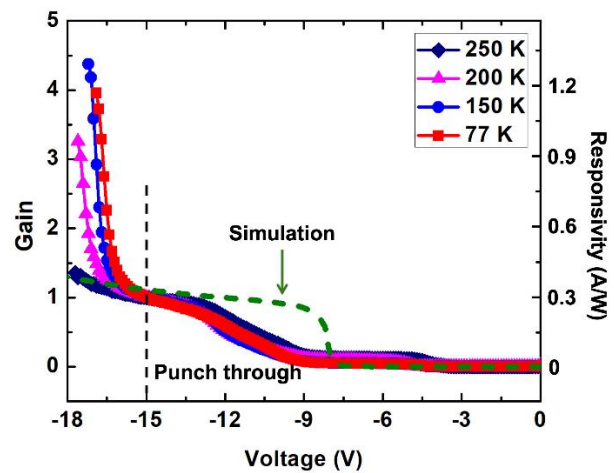

**Fig. 6.** Internal gain and responsivity as a function of voltage at different temperatures. The unity gain was taken at punch-through voltage of -15 V. Dashed line is simulation of gain at 250 K.

Figure 6 shows measured responsivity and gain as a function of voltage at different temperatures. Using identified punch-through voltage of -15 V from C-V measurement, the primary responsivity was measured as 0.28 A/W at 77 K, corresponding to the unity gain. The responsivity increases sharply to 1.12 A/W at 16.9 V (99% of breakdown voltage) with the internal gain of 4.0, due to the carrier multiplication process in Si layer. As temperature increases, the responsivity increasing rate reduces mainly because of decreased impact ionization rate and increased breakdown voltage at higher temperatures. The green dashed curve is simulated gain at 250 K, which shows agreement with measured data at low voltage and above punch-through. In addition, a device with mesa side length of 1000 μm was also characterized using the same method, exhibiting the primary responsivity of 0.07 A/W and maximum measured gain of 4.5 at 77 K, respectively. The temperature-dependent gain and responsivity of two devices are summarized in Table I.

Table I Summary of gain and responsivity

| Temperature (K) | Maximum measured gain | | Responsivity @ -16.9 V (A/W) | |
|---|---|---|---|---|
| | 500 μm | 1000 μm | 500 μm | 1000 μm |
| 77 | 4.0 | 4.5 | 1.12 | 0.30 |
| 150 | 4.4 | 3.8 | 0.72 | 0.28 |
| 200 | 3.3 | 1.8 | 0.26 | 0.13 |
| 250 | 1.4 | 1.2 | 0.22 | 0.10 |

It is worth noting that although a thin Ge buffer facilitates the electron transporting from absorber to Si multiplication layer, on the other hand it limits the Sn incorporation in GeSn absorber in order to keep low density of defects at Ge/GeSn interface and threading dislocations in GeSn. For higher Sn absorber, the buffer layer needs to be redesigned to tradeoff between carrier collection and material quality. Moreover, to improve the responsivity and gain, the reduction of dark current is needed. This can be achieved by surface passivation process to decrease the surface recombination. There might be other current leakage channel in present devices. More study will be performed to pinpoint the leakage mechanism to further improve the performance.

We would also like to point out that the electric filed simulation results show that the GeSn layer is fully depleted at -13 V, and the measured dark current increases steeply at -12 V (Fig. 3). This suggests that the actual punch through voltage could be around -13 V. In this study we conservatively selected -15 V as punch through, which may result in underestimated internal gain. If the -13 V punch through is selected, the gain of the 500 μm device will be calculated as 5.1 (verse 4.0) at 77 K, showing a 27.5% increase.

## V. CONCLUSION

The GeSn on Si SACM APDs were demonstrated. A thin Ge buffer layer of 100 nm design was applied to efficiently collect photo generated electrons. The spectral response shows the cut-off at 1850 nm, attributed to GeSn absorption. The punch-through voltage was identified at -15 V via C-V measurement. The primary responsivity was 0.28 A/W at 77 K, and the maximum measured multiplication gain of 4.5 was obtained.

## ACKNOWLEDGMENT

Shui-Qing Yu would like to thank Northrop Grumman Corporation (NGC) for gift funding to partially support this work. Many fruitful discussions and encouragements from Drs. Alex Toulouse and Leye Aina at NGC for this work are greatly appreciated.

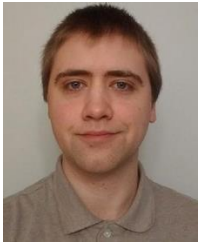

**Justin Rudie** received his B.S. in electrical engineering from Minnesota State University, Mankato in 2019. After a REU experience at the University of Arkansas he continued his studies there and received his M.S. in Material Engineering in 2022. He is a two-time recipient of the DoD's SMART Scholarship and will pursue a career at the Naval Surface Research Center in Crane, IN after completing his doctoral studies. His research focuses are in mid-IR imaging technology.

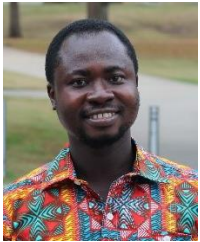

**Sylvester Amoah** received his B.S. degree in Physics from the Kwame Nkrumah University of Science and Technology, Kumasi, Ghana, in 2011 and an MS Degree in Physics from Western Illinois University, Macomb, Illinois, USA in 2018. Currently, he is doing his Ph.D. in electrical engineering at the Department of Electrical Engineering and Computer Science at the University of Arkansas, Fayetteville, USA.

His research interests include i) Novel Group-IV photonic materials such as SiGeSn materials, lasers and photodetectors, and ii) Novel platforms for photonics integrated circuits, particularly, Si photonics. He has published 15 articles in refereed journals and 8 articles in conference proceedings. He has also given/contributed presentations and invited talks at prestigious conferences such as OSA, CLEO and SPIE Photonics West.

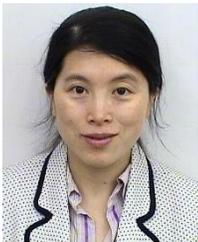

**Xiaoxin Wang** received the B.S. degree in 2000 and M. S. degree in 2002 in materials science and engineering from Tsinghua University, Beijing, China, and the Ph.D. degree in solid state electronics from the Institute of Semiconductors, Chinese Academy of Sciences, Beijing, in 2006. She is currently a Research Scientist at the Thayer School of Engineering, Dartmouth College, Hannover, NH, USA. She has been conducting research on photonic materials and devices, including phosphor materials, Si nanocrystal and nanowire light emitters, infrared and UV photodetectors, Ge/GeSn optoelectronic, solar-thermal coatings and visible laser lighting and communication. She has authored or coauthored more than 50 publications and awarded several U.S. patents. Dr. Wang is a member of the Materials Research Society, IEEE, and Optica.

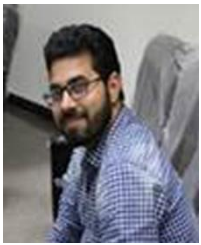

**Rajesh Kumar** did his Ph.D. in Physics on Material and Devices for Photothermal Applications from IIT Jodhpur, Rajasthan, India. He did his masters in Nano Science and Technology for GGS Indraprastha University, Delhi, India and Bachelor's in Electronics and Communication Engineering. During Ph.D. he visited NTU Taiwan under the TEEP program and worked on the synthesis of photoluminescence materials. After receiving his Ph.D., he joined NYCU Hsinchu Taiwan as a postdoc and worked on AlGaInP and GaN based Light-Emitting Diodes (LED). Then he joined the University of Arkansas, Fayetteville, USA as a postdoc and currently is working on InGaAs QW based lasers and GeSn based photodetectors.

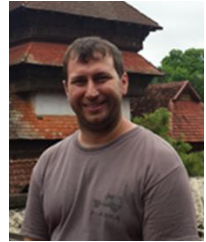

**Grey Abernathy** received a B.S. degree in Computer Engineering from the University of Arkansas in 2007, an M.S. in Nanoengineering from the Joint School of Nanoscience and Nanoengineering in 2016, and successfully defended his Ph.D. research for 2024 graduation in Microelectronics – Photonics from the University of Arkansas. He was a high school mathematics and robotics instructor at multiple Arkansas school districts from 2011 – 2014 and has been employed at Arktonics from 2021 – 2024. His research interests include Silicon Photonics, Group – IV emitters, and other materials science. He has published 15 journal articles and seven conference publications.

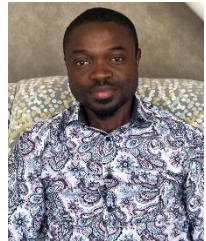

**Steven Akwabli** earned his B.S. degree in Physics from Kwame Nkrumah University of Science and Technology in Ghana in 2011, and an M.S. degree in Physics from Western Illinois University in Illinois in 2017. From 2017 to 2019, he taught Physics and Astronomy at the City Colleges of Chicago, Oakton Community College in Des Plaines, and Harper College in Palatine, Illinois as an adjunct. He then held an assistant professorship position at Clevelandstate Community College in Tennessee until 2022. Currently, he is pursuing his PhD in Electrical Engineering at the University of Arkansas, where he also serves as a graduate research assistant.

His research focuses on the growth and characterization of GeSn and SiGeSn materials, optoelectronic device fabrication and characterization, and the transport properties of GeSn quantum wells. He has shared his research findings at prestigious conferences such as the MIOMD and EFRC, contributing valuable insights into his field of study.

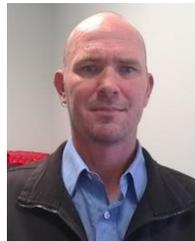

**Perry C. Grant** received a B.S. degree in Engineering Physics from Southern Arkansas University, Magnolia, AR in 2011, M.S (2013) and Ph.D. in Microelectronics-Photonics from the University of Arkansas in 2018. He was an Electrical Engineer with Applied Technology Associates to support ongoing projects with Air Force Research Laboratory Space Vehicles Directorate at Kirland AFB (2018-2021) and joined Air Force Research Laboratory Space Vehicles Directorate at Kirtland AFB as Electronics Engineer (2021-2024). He is currently a Research Scientist for Arktonics LLC.

His research interest includes i) Group IV alloys and devices for Infrared sensing and imaging, ii) Dissimilar material integration strategies for monolithic EOIR devices in the near-, mid-, and far-infrared, and iii) Semiconductor growth technologies and processes such as MBE and CVD. He has published 36 articles in refereed journals and 20 articles in conference proceedings. He has given/contributed to presentations and invited talks at CLEO, NAMBE, and IEEE meetings.

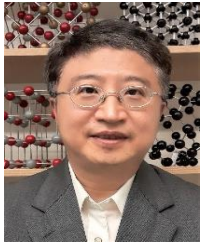

**Jifeng Liu** (Senior Member, IEEE) received the B.S. and M.S. degrees from Tsinghua University, Beijing, China, and the Ph.D. degree from the Massachusetts Institute of Technology, Cambridge, MA, USA, both in material science and engineering. He is currently a Professor and Program Area Lead of Materials Science and Engineering at Thayer School of Engineering, Dartmouth College, Hanover, NH, USA. His major research field is photonic materials and devices for communication, sensing, and energy applications. He has authored or coauthored >100 peer-reviewed journal papers, >80 conferences papers, and 7 book chapters, which have been cited >13,500 times according to Google Scholar. He has also been granted 18 U.S. patents. Prof. Liu is a Fellow of Optica. He served as a General Co-Chair of 2023 IEEE Silicon Photonics Conference, and a Science & Innovation Program Co-Chair of 2024 Conference of Lasers and Electro-Optics (CLEO).

**Baohua Li** received her B.S. and M.S. degrees in mathematics from Sichuan University, Chengdu, China, in 1999 and 2002, respectively, and the Ph.D. degree from the Department of Electrical Engineering from Arizona State University, Tempe, in 2009. She is currently with Arktonics, LLC. Her current research interests include SiGeSn material growth and device development.

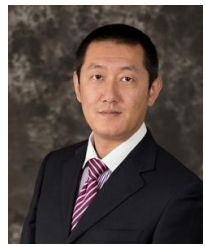

**Wei Du** received the B.S. degree in Physics from Peking University, Beijing, China, in 2003, and Ph.D degree in Electrical Engineering from Institute of Semiconductors, Chinese Academy of Sciences in 2008. He was a postdoctoral research associate at several institutes including Arizona State University, Shizuoka University (Japan), Washington State University, and University of Arkansas from 2008-2017. In 2017, he joined the Department of Electrical Engineering and Physics at Wilkes University, Wilkes-Barre, PA, USA as an assistant professor, where he received the Outstanding New Faculty Award in 2019, the President's Award for Excellence in Scholarship in 2020, and the Outstanding Advisor Award in 2022. Currently, he is an Associate Professor of the Department of Electrical Engineering and Computer Science at the University of Arkansas.

His research interests include i) Novel Group-IV photonic materials such as SiGeSn materials and devices, ii) Novel platforms for photonics integrated circuits including Si photonics and Sapphire, and iii) Nanophotonics devices such as photonic crystal lasers. He has published more than 100 articles in refereed journals and 150 articles in conference proceedings. He has also given/contributed presentations and invited talks at prestigious conferences such as CLEO, SPIE Photonics West, and IEEE Photonics Conference. He chaired numerous conference sections including IEEE Summer Topic, IEEE Si photonics Conference, etc.

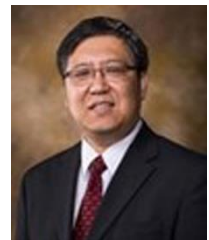

**Shui-Qing (Fisher) Yu** received the B.S. and M.S. degrees in Electronics from Peking University, Beijing, China, in 1997 and 2000, respectively, and Ph. D degree in Electrical Engineering from Arizona State University (ASU) in 2005. He was a postdoctoral research associate and an assistant research professor at ASU from 2005-2008. In 2008, he joined the Electrical Engineering Department, University of Arkansas (UA), Fayetteville, AR, USA as an assistant professor. Currently, he is a Professor of the Department of Electrical Engineering and Computer Science at UA and holds a Twenty-First Century Research Leadership Chair. He also directs an Energy Frontier Research Center "Manipulation of Atomic Ordering for Manufacturing Semiconductors (μ-ATOMS)" funded by the Department of Energy.

His research interests include i) Novel electric and photonic materials such as SiGeSn(Pb) materials and devices, Dissimilar material integration, and Low dimensional quantum materials and ii) Novel optoelectronics future integrated photonics such as Alternative platform for integrated photonics, Novel optoelectronics using nanostructures and new phenomena, and High temperature optoelectronics and its applications for power electronics. In 2012, he received the Faculty Early Career Development award from the National Science Foundation.

He has published 161 articles in refereed journals and 209 articles in conference proceedings. He has also given/contributed 217 presentations and 51 invited talks and holds 13 patents. He is a fellow of Optica and a Senior Member of IEEE and SPIE. He chaired numerous conferences including "Mid-IR Optoelectronics: Materials and Devices (MIOMD)" in 2018 and 2023.